\documentclass[conference,letterpaper]{IEEEtran}
\usepackage[letterpaper, left=54pt, right=54pt, bottom=54pt, top = 54pt]{geometry}
\newgeometry{left=54pt, right=54pt, bottom=54pt, top = 72pt}
\IEEEoverridecommandlockouts

\usepackage{amsmath}
\usepackage{amsfonts}
\usepackage{hyperref}
\usepackage{subcaption}

\usepackage{tikz}
\usepackage{xfrac}
\usepackage{siunitx}
\usepackage{float}
\usepackage{pgfplots}
\usepackage{tikz-3dplot}
\usetikzlibrary{calc}
\pgfplotsset{width=7cm,compat=1.8}

\usepackage{color}
\usepackage{soul}

\usepackage[style=ieee, 
citestyle=numeric-comp,
sorting=none]{biblatex}
\def\BibTeX{{\rm B\kern-.05em{\sc i\kern-.025em b}\kern-.08em
    T\kern-.1667em\lower.7ex\hbox{E}\kern-.125emX}}
\begin{document}

\title{Overtaking Maneuvers on a Nonplanar Racetrack\\
\thanks{A video can be found at \href{https://youtu.be/5S-VfrV-d28}{https://youtu.be/5S-VfrV-d28}}
\thanks{Source code: \href{https://github.com/thomasfork/Nonplanar-Vehicle-Control}{https://github.com/thomasfork/Nonplanar-Vehicle-Control}}
\thanks{\textsuperscript{1}Thomas Fork and Francesco Borrelli are with the Department of Mechanical Engineering, University of California, Berkeley, USA \texttt{\{fork, fborrelli\}@berkeley.edu}}
\thanks{\textsuperscript{2}H. Eric Tseng is with the Ford Motor Company}
}

\author{\IEEEauthorblockN{Thomas Fork\textsuperscript{1} }
\and
\IEEEauthorblockN{H. Eric Tseng\textsuperscript{2} }
\and
\IEEEauthorblockN{Francesco Borrelli\textsuperscript{1} }
}

\maketitle

\begin{abstract}
We leverage game theory and a new vehicle modeling approach to compute overtaking maneuvers for racecars on a nonplanar surface. We solve for equilibria between noncooperative racing agents and demonstrate that with the novel nonplanar vehicle dynamics, overtaking can be achieved in situations where simpler models cannot provide a winning strategy. 
\end{abstract}

%%%%%%%%%%%%%%%%%%%%%%%%%%%%%%%%%%%%%%%%%%%%%%%%%%%%%
\section{Introduction}
Game-theoretic approaches to racing are widespread \cite{8643396, 9439814, 9329208} and a key focus of noncooperative vehicle control. However, these approaches are based on vehicle models that assume the road surface is flat, which is seldom the case for real raceways such as Figure \ref{fig:tri-oval}. We leverage recent advancements in nonplanar control-oriented vehicle models \cite{fork2021models} to study this problem, in particular the task of finding a collision-free overtaking maneuver, which we do in an offline setting. Our contributions are twofold: First we adapt a standard raceline optimization problem for an overtaking problem. Second, we use this problem and our nonplanar vehicle modeling approach to generate overtaking maneuvers on a nonplanar racetrack. We demonstrate that with nonplanar vehicle models, feasible overtaking trajectories can be found where not present otherwise. This paper is structured as follows: In section \ref{sec:background} we introduce background on our modeling approach and comment on the specific vehicle models used in section \ref{sec:models}. We introduce a trajectory planning approach in section \ref{sec:planning} which we adapt for overtaking in section \ref{sec:overtaking}. We present results of our approach in section \ref{sec:results} and conclude in section \ref{sec:conclusion}.

\section{Background} \label{sec:background}
\begin{figure}
    \centering
    \includegraphics[width=\linewidth]{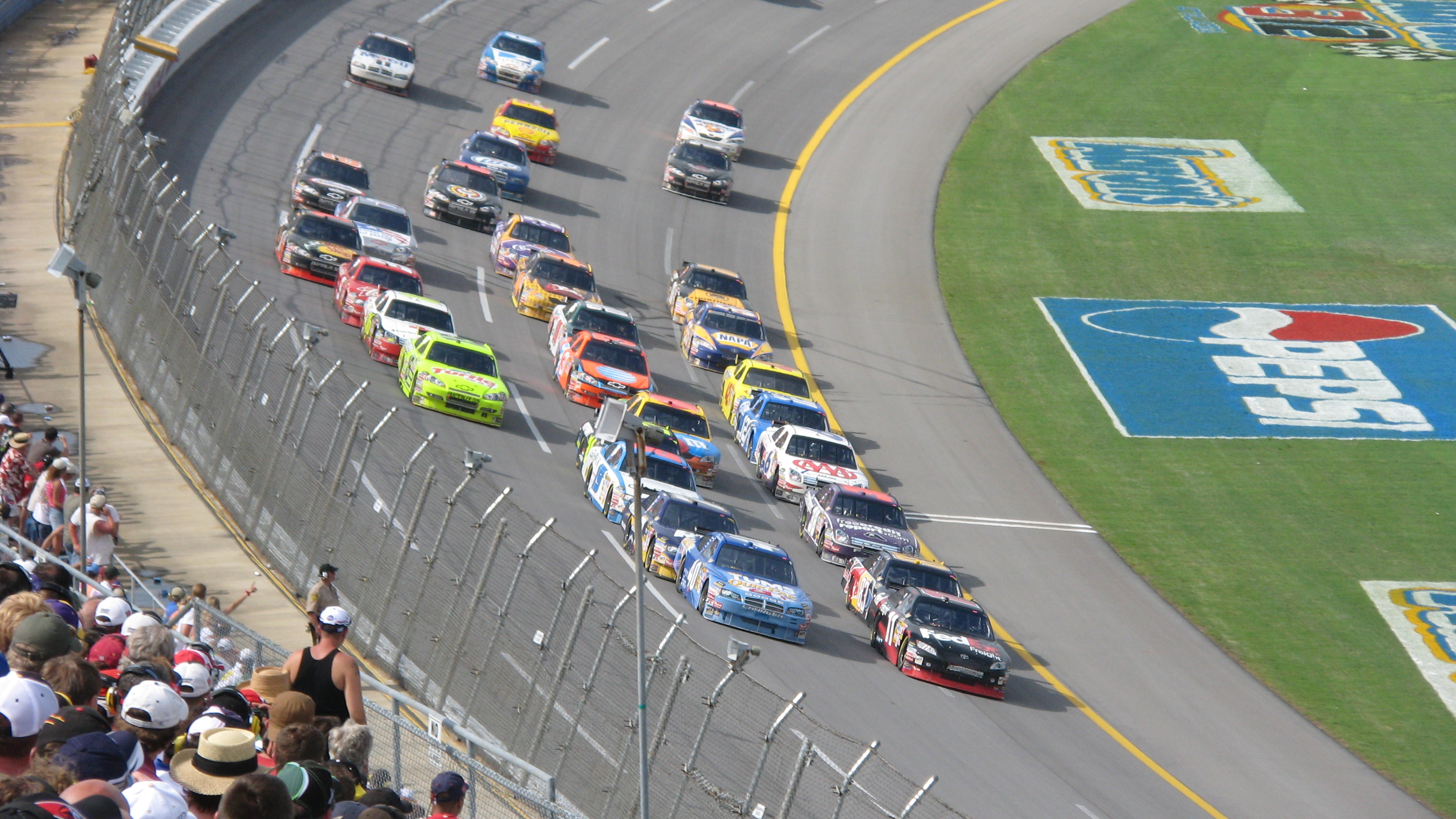}
    \caption{Tri-Oval turn at the Talladega Superspeedway.}
    \label{fig:tri-oval}
    \small Image "Through the tri-oval" by Curtis Palmer, licensed under \href{https://creativecommons.org/licenses/by/2.0/deed.en}{Creative Commons Attribution 2.0 Generic}
\end{figure}

We build on our work in \cite{fork2021models} and \cite{fork2022models} which introduced control-oriented vehicle models for nonplanar road surfaces. Our models treat a vehicle as a rigid body in tangent contact with a nonplanar road surface, which is a reasonable approximation of real racecar motion such as Figure \ref{fig:tri-oval}. Unlike previous nonplanar vehicle models \cite{3d_part_1}, our approach enforces tangent contact exactly and can be applied to arbitrary road surfaces, such as roads with curved cross-section. We represent the road surface with a parametric surface and use the surface parameterization to describe vehicle pose. By adding rigid body dynamics and vehicle motion constraints we obtain numerically tractable vehicle models on an arbitrary road surface. Though more general in scope, these models follow the same format as existing control-oriented vehicle models, such as \cite{7225830}. We refer the unfamiliar reader to \cite{fork2022models}, wherein a dynamic vehicle model was introduced and used for computing racelines on a nonplanar road surface, as we build on this problem to generate overtaking maneuvers. 

\section{Vehicle Models} \label{sec:models}
We use two nonplanar vehicle models from \cite{fork2022models}. First we use the two-track dynamic vehicle model which considers dynamic weight distribution and uses a combined slip Pacejka model for each tire. We also use the nonplanar kinematic bicycle model, which only considers net normal force on the vehicle and incorporates a friction cone constraint for path planning. While the two-track model constrains the normal force on each tire, the kinematic bicycle constrains the net normal force on the vehicle in a conservative manner as justified in \cite{fork2022models}. 

In \cite{fork2022models} we demonstrated that on flat roads the kinematic and two-track models can achieve similar results, but the two-track model excels on suitable nonplanar geometry. We build on this observation to use the two-track model for an agent planning an overtaking maneuver. We do not use the planar kinematic bicycle tested in \cite{fork2022models}, as it was shown to be inferior to the nonplanar kinematic bicycle, which we do use. 

\begin{figure}
    \centering
    \includegraphics[width = 0.95\linewidth]{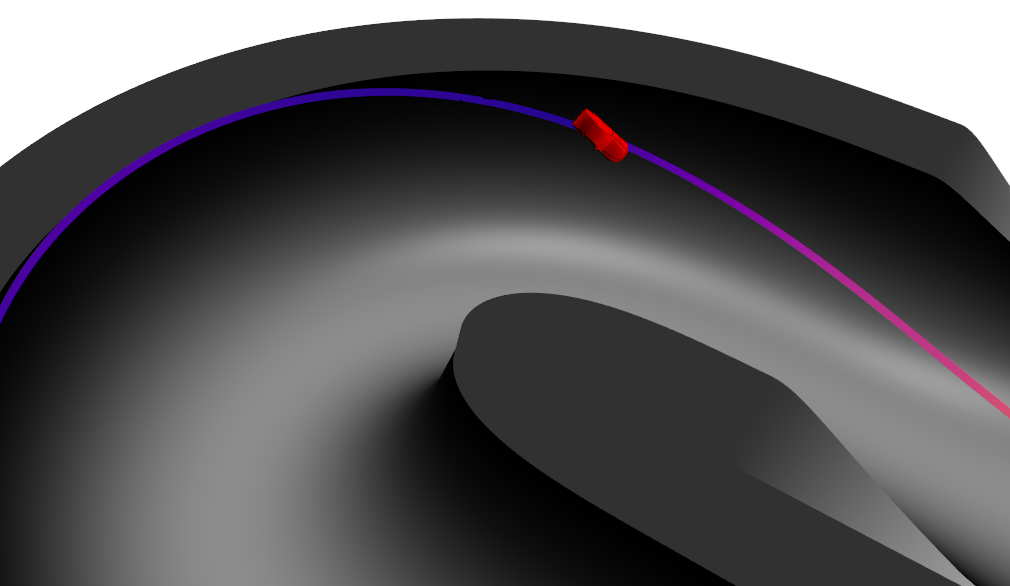}
    \caption{An example of a nonplanar single vehicle raceline with a two-track vehicle model.}
    \label{fig:example}
\end{figure}

\section{Trajectory Planning Problem} \label{sec:planning}
We briefly recap the raceline problem of \cite{fork2022models} before repurposing it for planning an overtaking maneuver. In this setting, our goal is to minimize the time it takes a vehicle to reach one point on a racetrack from another, for instance reaching the finish line from a point 300m behind it and around a turn. This problem is solved by discretizing the remaining distance into intervals of uniform length and enforcing a vehicle model using a direct collocation \cite[ch.~10]{biegler_book} \cite{christ2021time, 3d_part_2} to implement the continuous time vehicle model over the problem domain\footnote{Direct collocation may also be referred to as a pseudospectral method \cite{patterson2014gpops}}. Constraints are enforced to keep the vehicle within track boundaries, limit tire normal forces and enforce friction limits. We also apply a small regularization penalty to input and input rate. An example of a single vehicle raceline with the two-track vehicle model is shown in Figure \ref{fig:example}.

This problem setup is insufficient for the multi-vehicle case, where collision avoidance imposes new and interactive constraints between agents. This interaction is often modeled using game theory; we introduce such an approach next.

\section{Overtaking Problem} \label{sec:overtaking}
We adapt the raceline problem of \cite{fork2022models} for an overtaking problem. We consider a scenario where an Ego vehicle and a Target vehicle, initially at identical speed, are about to enter a curved turn. Both attempt to reach the end of the turn as quickly as possible, with the Ego vehicle attempting to overtake the Target vehicle. While the Ego vehicle uses the two-track vehicle model from \cite{fork2022models}, the Target vehicle uses a nonplanar kinematic bicycle model. As the Target vehicle is initially in front, the Ego vehicle must yield to its maneuvers. 

To capture this difference in models and right of way we set up a leader-follower game with imperfect knowledge as follows:
\begin{enumerate}
    \item The Target vehicle computes an optimal raceline around the upcoming turn, ignoring the Ego vehicle
    \item The Ego vehicle does the same, but only to obtain a warmstart trajectory
    \item The Ego vehicle solves the same problem once more, but with a collision avoidance constraint to avoid the Target vehicle's optimal trajectory
\end{enumerate}
As a result, the Target vehicle plans a route that is optimal according to its vehicle model and the Ego vehicle adapts accordingly, its only advantage being a more advanced vehicle model. Steps 1 and 2 above are identical to the raceline problem in \cite{fork2022models}; we provide further details on step 3. 

\subsection{Collision Avoidance Constraint}
We implement the collision avoidance constraint by approximating each vehicle as a circle and enforcing that each circle does not overlap the other. We chose this overconservative approach as it is simple to implement and did not negatively impact the scenarios we tested. We remark that exact approaches such as \cite{9062306} can be integrated in our approach as well. 

\begin{figure}
    \centering
    \includegraphics[width = 0.95\linewidth]{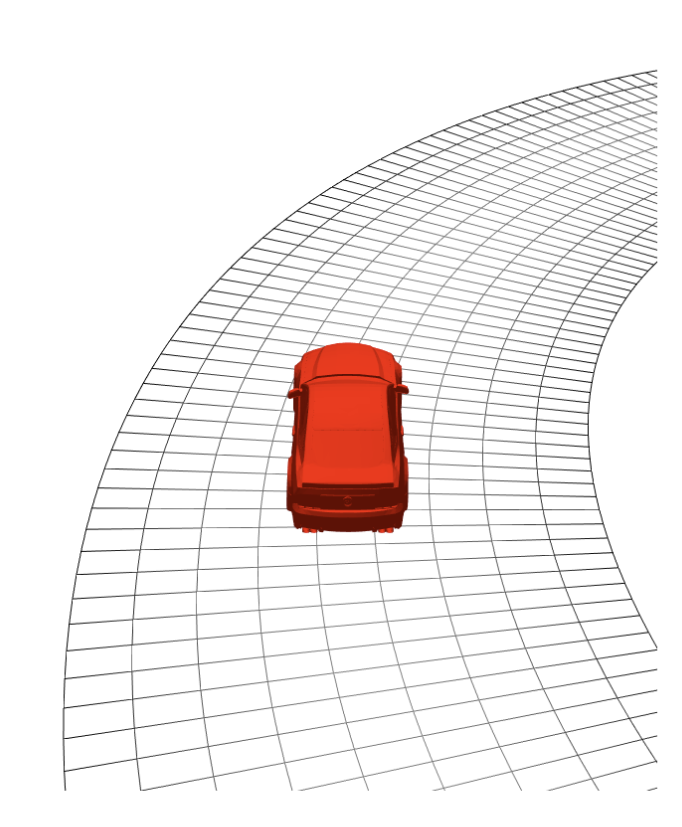}
    \caption{Frenet Frame contour lines, note the coordinates are compressed on the inside of the turn.}
    \label{fig:frenet}
\end{figure}

Two factors complicate the circular collision avoidance constraint. First, since our problem is expressed in the coordinate system of the parametric surface, information on the global, three-dimensional position of the vehicles is absent. As a result, if $\Delta_s$ and $\Delta_y$ are position differences between Ego and Target vehicles in each variable, enforcing $\sqrt{\Delta_s^2 + \Delta_y^2} \leq 2r$, where $r$ is the radius of each vehicle's outer circle, does not generally guarantee collision avoidance. This is a result of the coordinate system compressing where necessary for the surface parameterization, for instance on the inside of a turn in the Frenet frame, illustrated in Figure \ref{fig:frenet}. This is an issue only on the inside of the centerline; since we are free to make the centerline the inner edge of the track we do so where possible, avoiding potential collisions from this effect.

Second, The Ego vehicle trajectory is discretized in space, meaning the time to reach a state in the optimization problem is a variable itself. As a result, it is necessary to interpolate the Target vehicle trajectory for use as a constraint, since the relative position of the Target vehicle changes throughout the optimization problem. We do so using the interpolating polynomial implicit in our collocation method, wherein the state within a collocation interval has a polynomial form \cite[Eqn. 10.4]{biegler_book}. As a result, at any collocation point in the Ego vehicle problem we can compute the corresponding state of the Target vehicle by computing the time at the current point and interpolating the Target vehicle trajectory with that time. The time at a collocation point is the sum of the length of previous intervals and a fraction of the current interval that depends on where in the interval the point lies, allowing us to implement the interpolation in our optimization problem. 

\section{Results} \label{sec:results}

\begin{figure*}
    \centering
    \includegraphics[width = 0.95\linewidth]{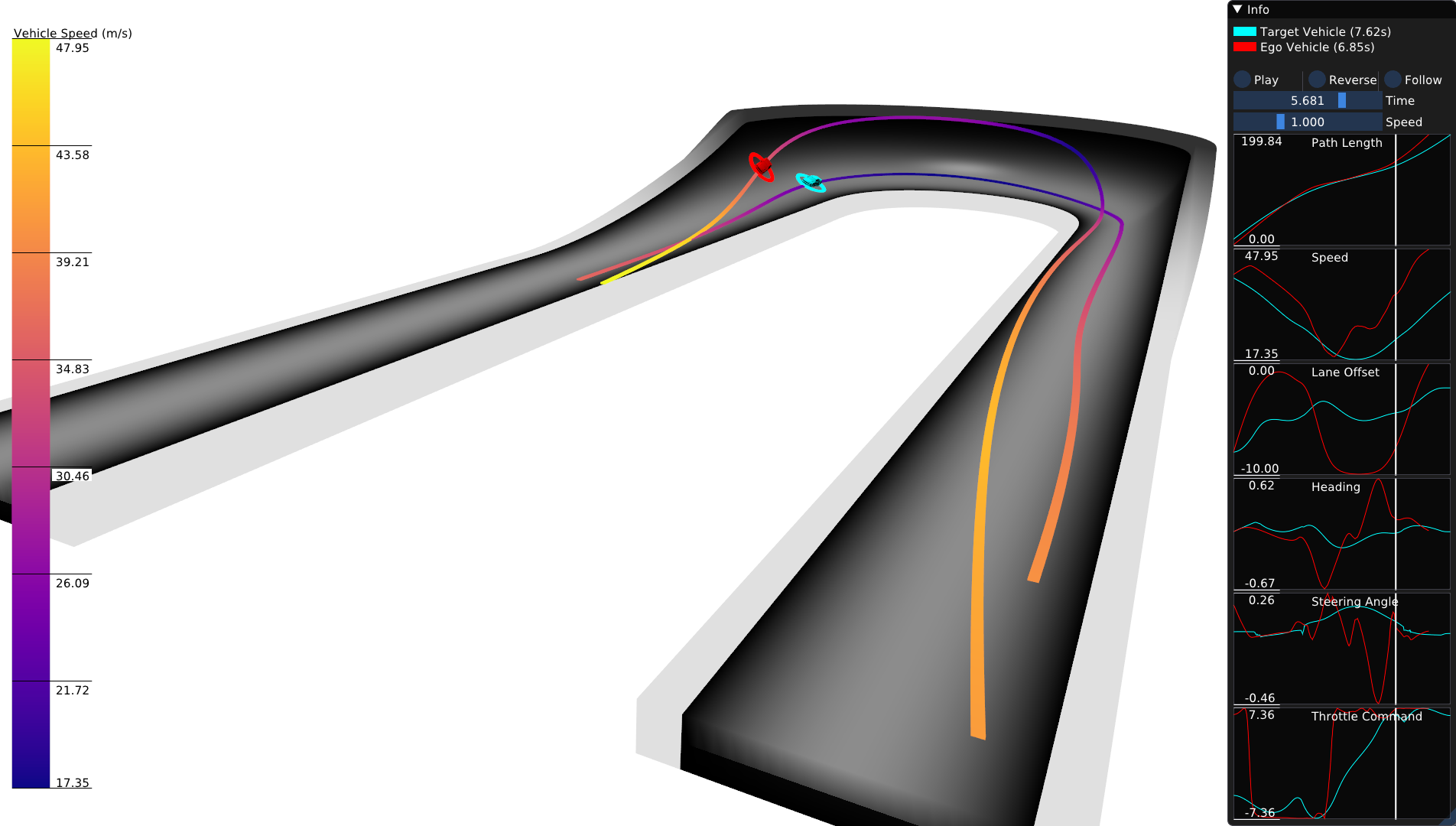}
    \caption{U-turn overtaking maneuver. Trajectories are colored by speed and vehicles are colored by agent. Agent color scheme and trajectory duration is provided in the legend. (Video: \href{https://youtu.be/5S-VfrV-d28}{https://youtu.be/5S-VfrV-d28})}
    \label{fig:overtaking_maneuver}
\end{figure*}

\begin{figure*}[h]
\begin{subfigure}{.33\linewidth}
  \centering
  \includegraphics[height=4cm]{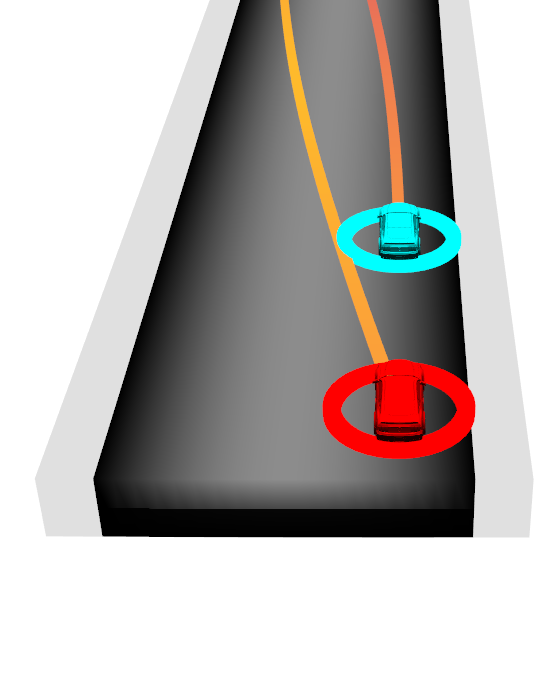}
  \caption{Beginning of the maneuver.}
  \label{fig:sub1}
\end{subfigure}%
\begin{subfigure}{.33\linewidth}
  \centering
  \includegraphics[height=4cm]{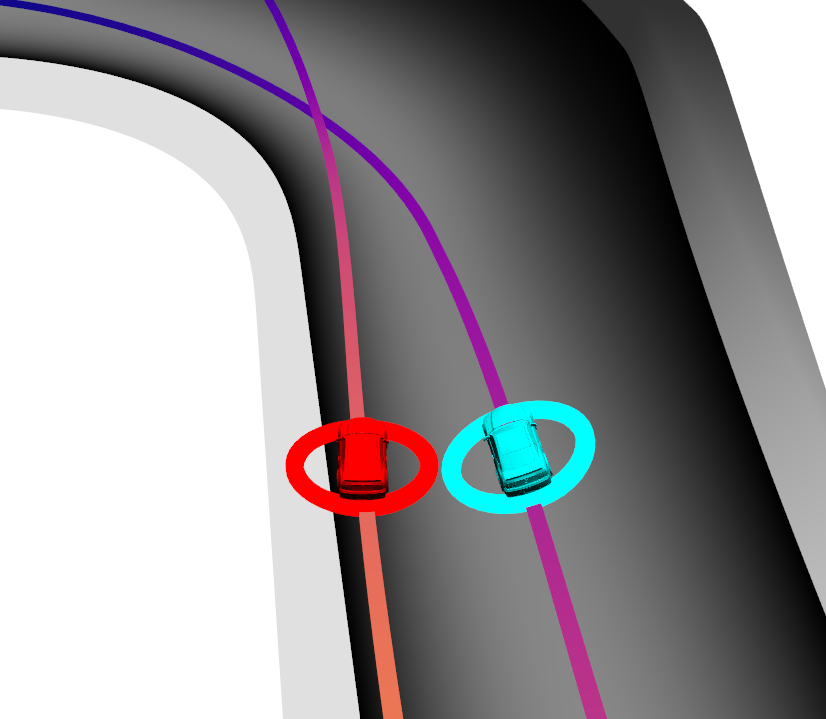}
  \caption{Passing the Target vehicle.}
  \label{fig:sub2}
\end{subfigure}%
\begin{subfigure}{.33\linewidth}
  \centering
  \includegraphics[height=4cm]{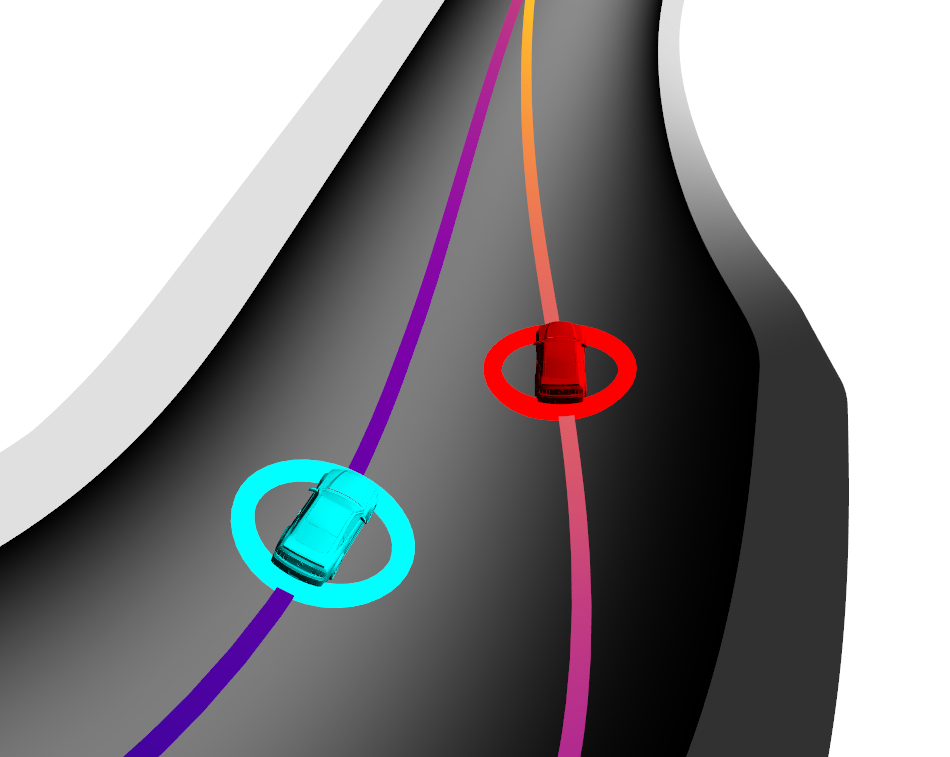}
  \caption{Ego vehicle pulling ahead out of the turn.}
  \label{fig:sub3}
\end{subfigure}%
\caption{Key points during the U-turn overtaking maneuver, circles represent collision buffers.}
\label{fig:overtaking_details}
\end{figure*}

We set up the described overtaking problem for two scenarios: a U-turn wherein the profile of the turn is an arc segment and a banked chicane. For the chicane the collision avoidance constraint is not guaranteed to be overconservative as the vehicles may travel on the inside of the centerline during both turns. However, this did not result in collisions during experiments. We implemented all vehicle models, the previously described surface parameterization and the raceline problem symbolically in CasADi~\cite{Andersson2018}, which was then solved using IPOPT~\cite{ipopt} with the linear solver MUMPS~\cite{MUMPS_1}. All programs were run and timed on an AMD Ryzen 5700U CPU at 4.3GHz. For the U-turn scenario we started both vehicles at a speed of $40$ meters per second with the Target vehicle $10$ meters directly ahead of the Ego vehicle. For the chicane scenario we used speeds of $10$ meters per second and the Target vehicle $5$ meters in front due to the tighter turns. We chose these initial configurations due to their difficulty: when both vehicles use the same model, the vehicle in front can simply follow its optimal raceline to avoid being overtaken. Where not stated otherwise, the Target vehicle used a nonplanar kinematic bicycle model and the Ego vehicle used a nonplanar two-track vehicle model. These models were used and compared previously in \cite{fork2022models}. 

\begin{figure}
    \centering
    \includegraphics[width = 0.7\linewidth]{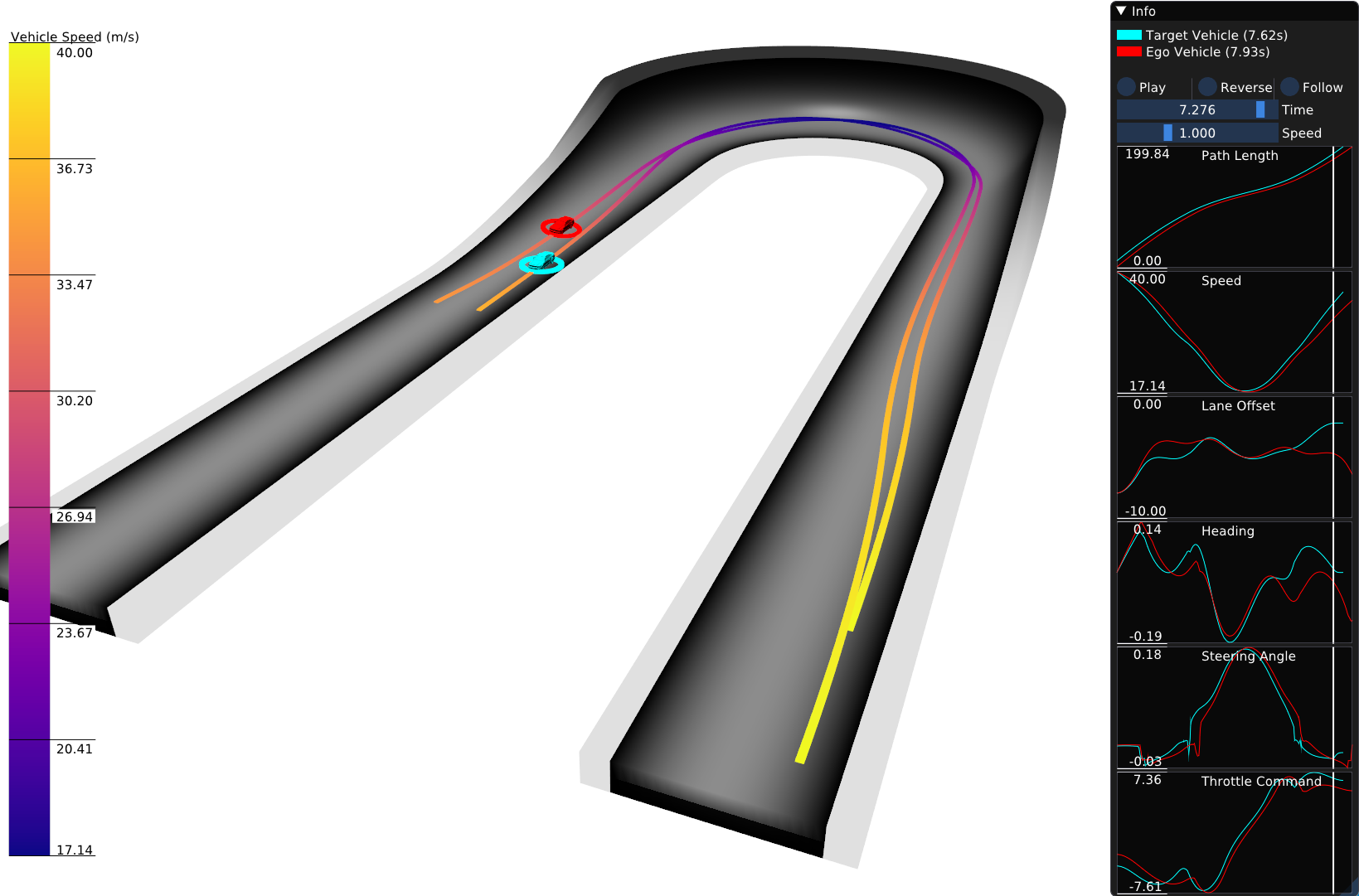}
    \caption{Failure to overtake on the U-turn when both agents use the kinematic vehicle model.}
    \label{fig:failed_overtaking}
\end{figure}

First, using the U-turn scenario, we solve our overtaking problem with both Target and Ego vehicle using the kinematic bicycle model (Figure \ref{fig:failed_overtaking}). As expected, the Ego vehicle is unable to overtake the Target vehicle, since the Target vehicle's route is optimal and leaves no room for overtaking. However, with the two-track vehicle model, the Ego vehicle is able to overtake when the Target vehicle slows down for the turn. The Ego vehicle is then able to complete the turn by exploiting nonplanar geometry to maintain a higher speed and exit the turn ahead of the Ego vehicle. The two overall trajectories are shown in Figure \ref{fig:overtaking_maneuver} and details of several points of the maneuver are shown in Figure \ref{fig:overtaking_details}. It took approximately 4 seconds to set up and solve the Target vehicle raceline problem, 24 seconds to solve for the Ego vehicle warmstart and 2 minutes to solve for the overtaking maneuver (times include setting up the problem). A video of the resulting maneuver can be found at \href{https://youtu.be/5S-VfrV-d28}{https://youtu.be/5S-VfrV-d28}

\begin{figure*}
    \centering
    \includegraphics[width = 0.95\linewidth]{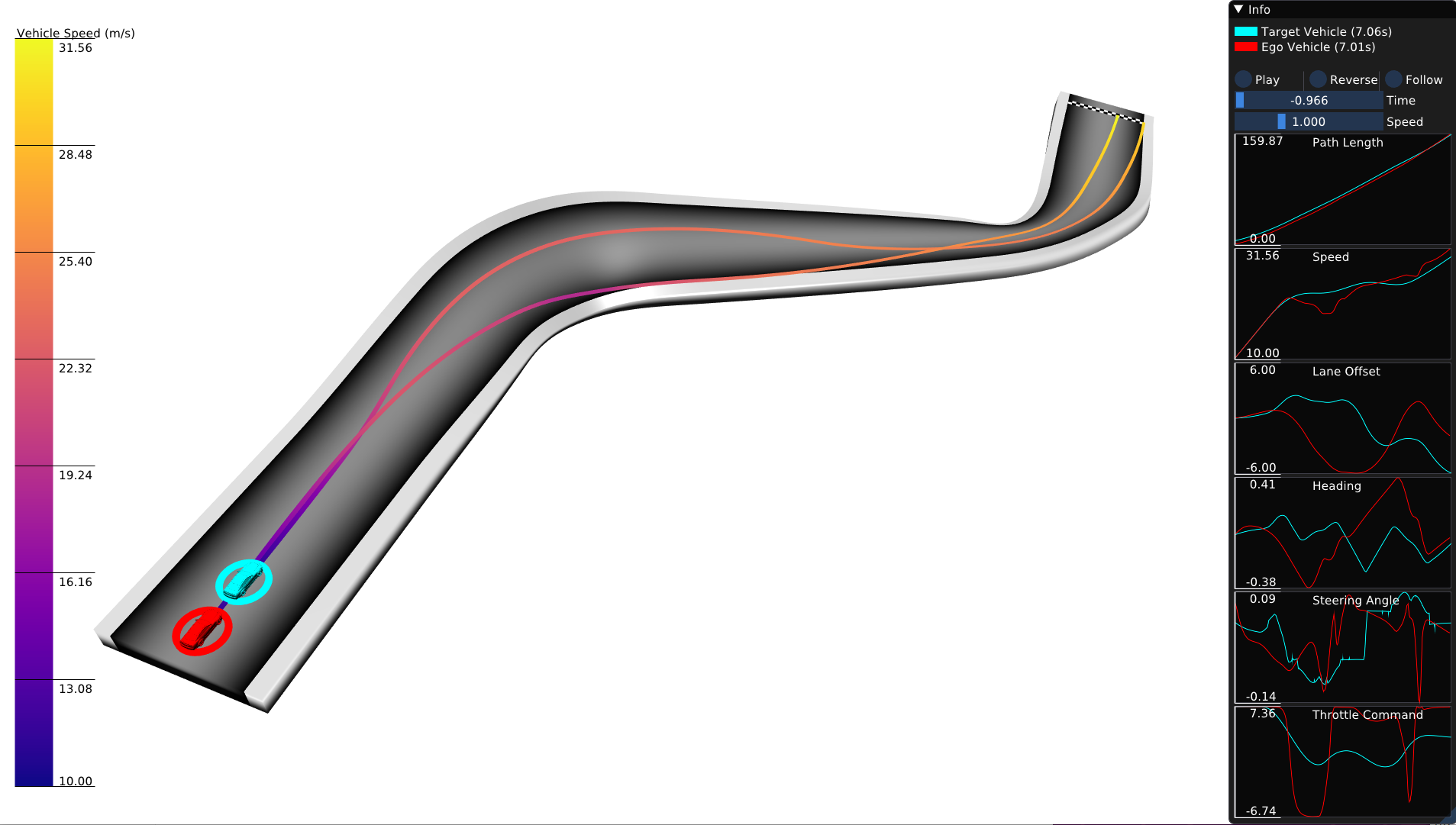}
    \caption{Overtaking maneuver on the banked chicane. Trajectories are colored by speed and vehicles are colored by agent. Agent color scheme and trajectory duration is provided in the legend.}
    \label{fig:chicane_overtaking_maneuver}
\end{figure*}

\begin{figure*}[h]
\begin{subfigure}{.33\linewidth}
  \centering
  \includegraphics[height=4cm]{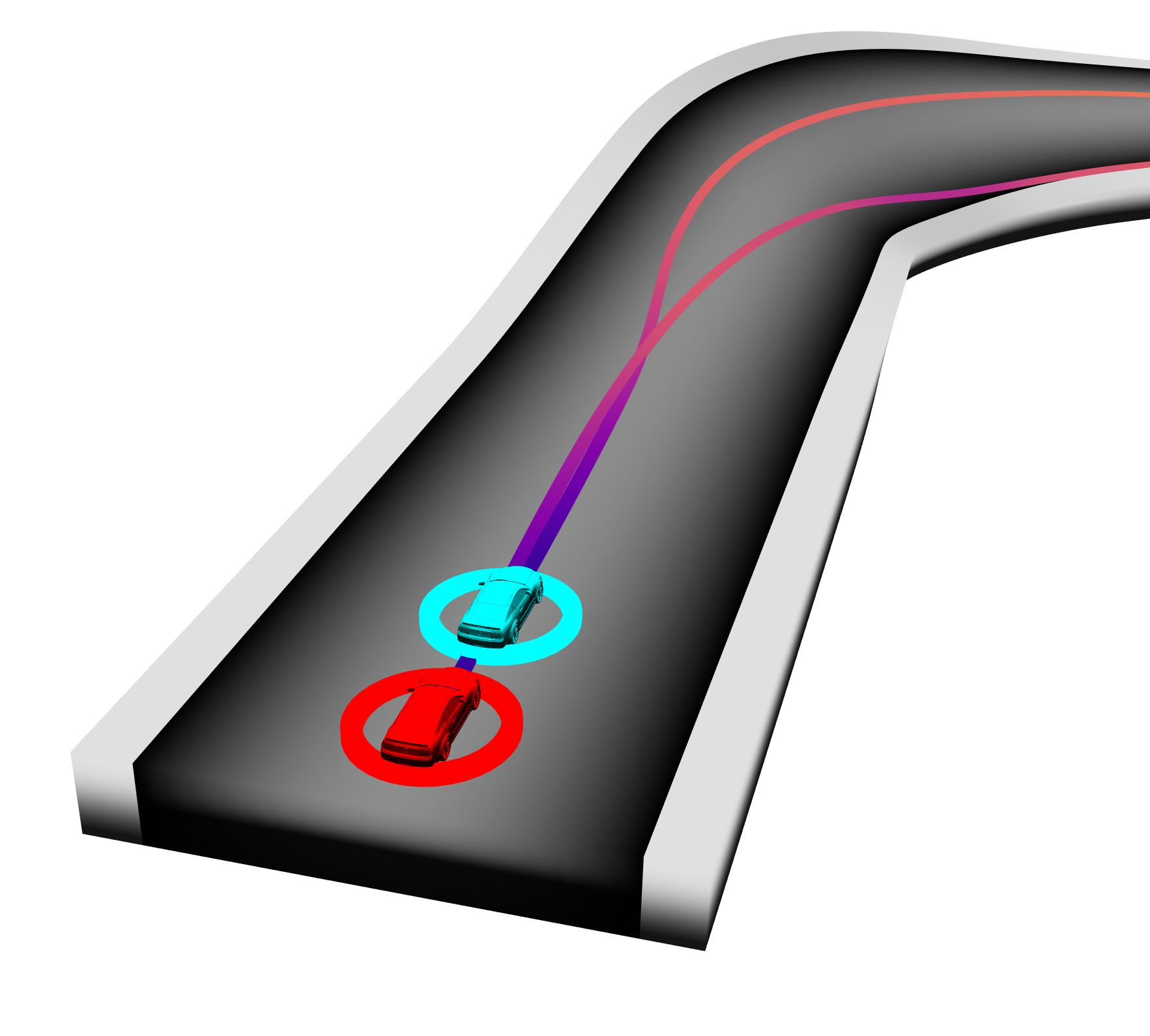}
  \caption{Beginning of the maneuver.}
  \label{fig:chicane_sub1}
\end{subfigure}%
\begin{subfigure}{.33\linewidth}
  \centering
  \includegraphics[height=4cm]{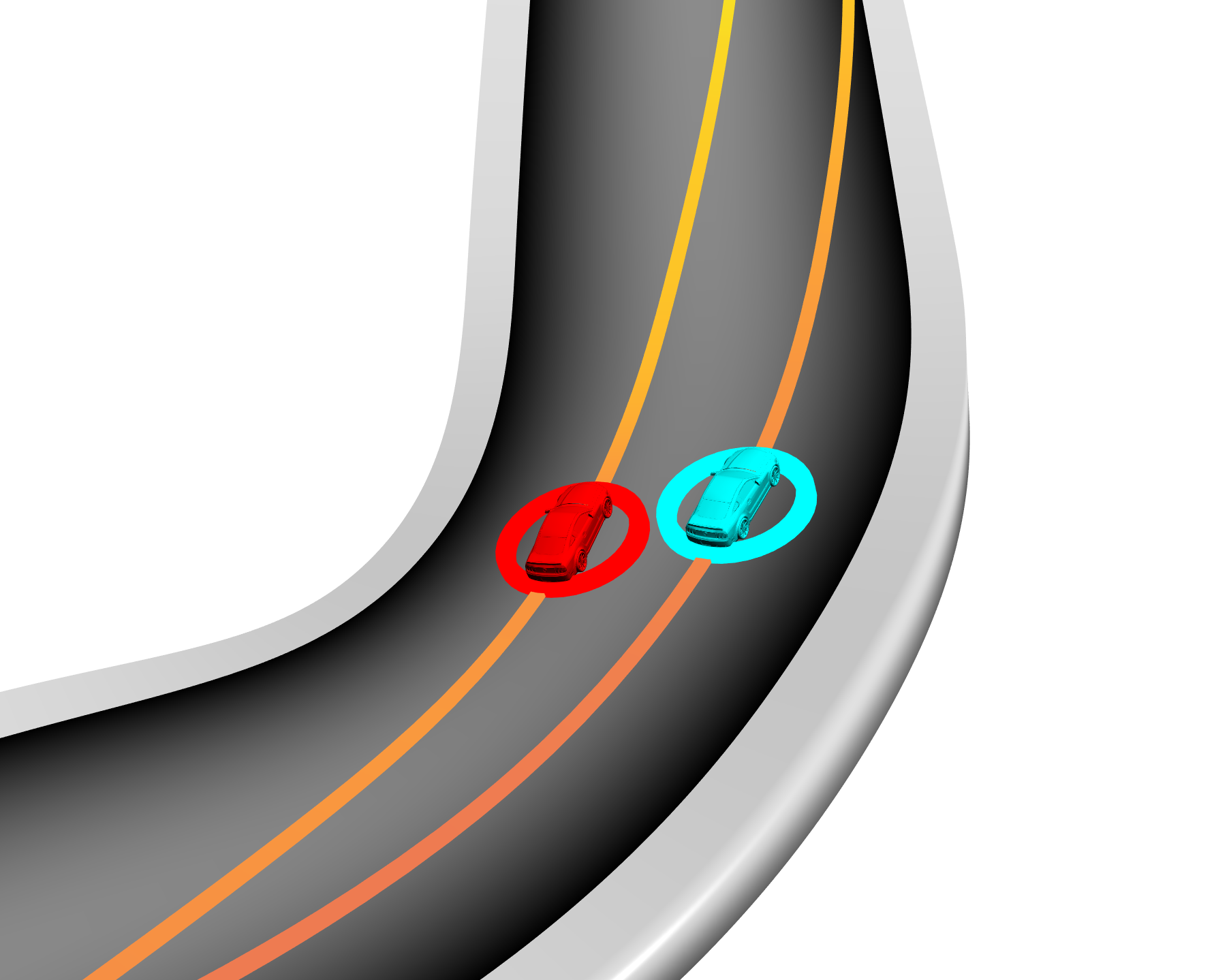}
  \caption{Passing the Target vehicle.}
  \label{fig:chicane_sub2}
\end{subfigure}%
\begin{subfigure}{.33\linewidth}
  \centering
  \includegraphics[height=4cm]{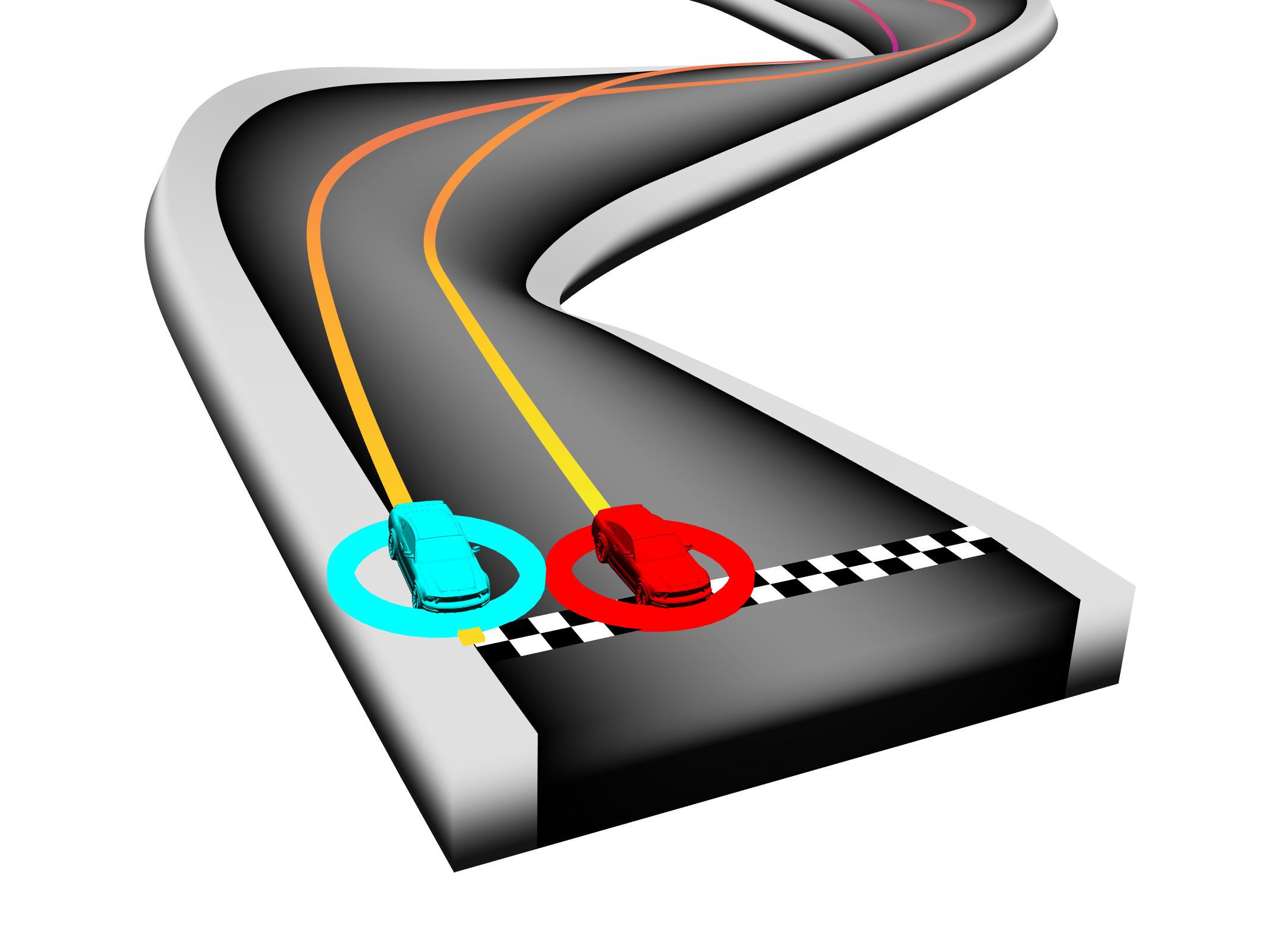}
  \caption{Ego vehicle reaching finish line just ahead of Target vehicle.}
  \label{fig:chicane_sub3}
\end{subfigure}%
\caption{Key points during the chicane overtaking maneuver, circles represent collision buffers.}
\label{fig:chicane_overtaking_details}
\end{figure*}

\begin{figure}
    \centering
    \includegraphics[width = 0.8\linewidth]{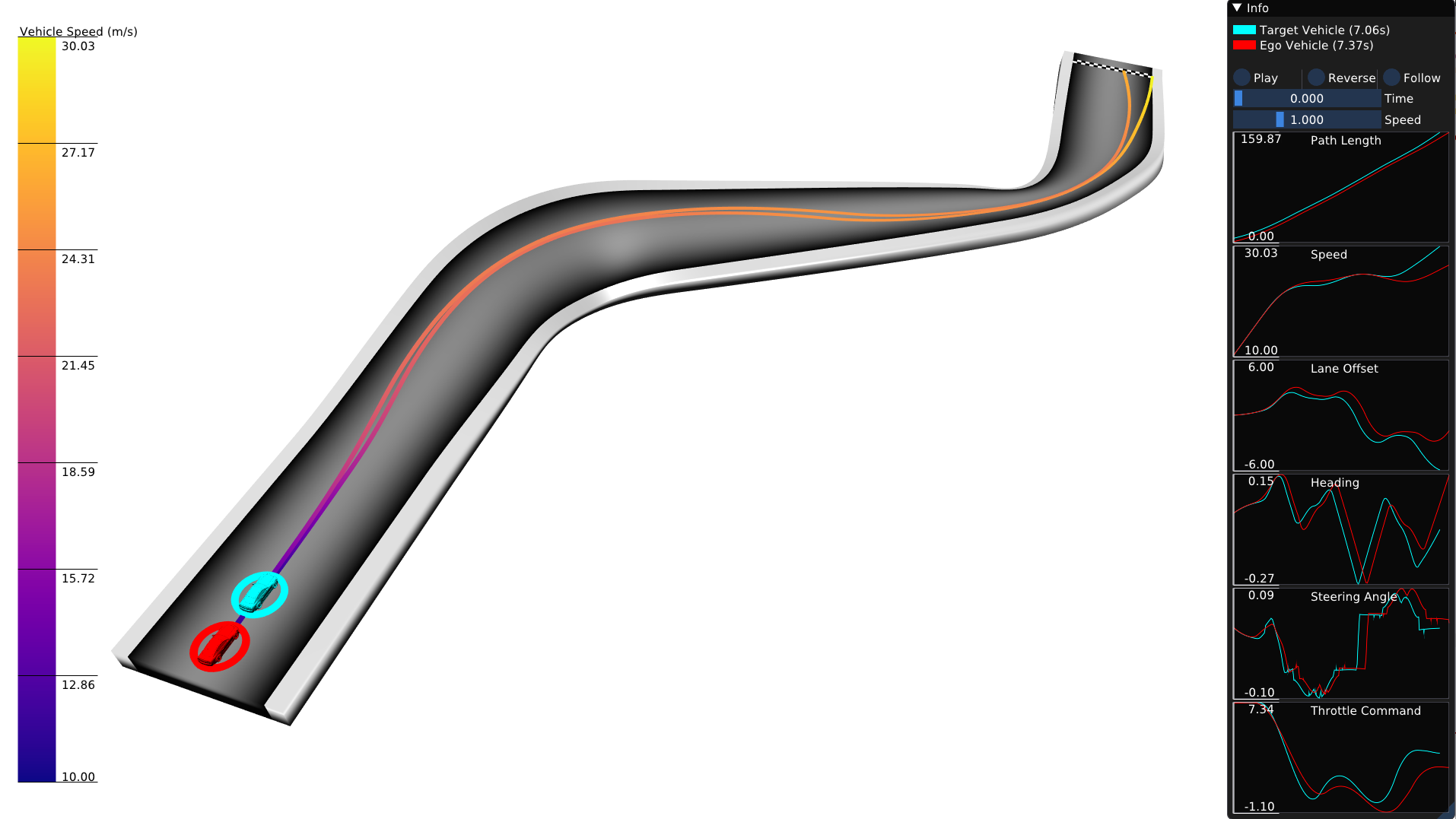}
    \caption{Failure to overtake on the chicane when both agents use the kinematic vehicle model.}
    \label{fig:chicane_failed_overtaking}
\end{figure}

Second we repeat the same experiments on the chicane scenario. As in the U turn scenario, when both vehicles use the same model the Ego vehicle is unable to overtake the Target vehicle, which starts in front and leaves no opportunity for the Ego vehicle to overtake (Figure \ref{fig:chicane_failed_overtaking}). However, when the Ego vehicle has the advantage of using the two track model it overtakes during the second bend of the chicane, and reaches the finish line just ahead of the Target vehicle, as shown in Figure \ref{fig:chicane_overtaking_maneuver}. Details on the beginning and end of the chicane maneuver, as well as the moment of overtaking the Target vehicle are highlighted in Figure \ref{fig:chicane_overtaking_details}.

\section{Conclusion} \label{sec:conclusion}
We presented an approach for computing overtaking maneuvers on a nonplanar road surface. We leveraged a new vehicle modeling approach to solve this problem in a tractable manner and demonstrated the potential for our method to find overtaking maneuvers when competing with less advanced models. We demonstrated that this is possible even when the Target vehicle performs optimally, so long as the Ego vehicle leverages suitable nonplanar geometry. 


\printbibliography
\end{document}